\documentclass[11pt]{article}

\usepackage[a4paper,margin=1in]{geometry}

\usepackage[T1]{fontenc}
\usepackage[utf8]{inputenc}
\usepackage{lmodern}
\usepackage{microtype}
\usepackage{amsmath,amssymb,amsfonts,mathtools}
\usepackage{graphicx}
\usepackage{booktabs}
\usepackage{csquotes}
\usepackage{hyperref}
\usepackage[capitalise]{cleveref}
\usepackage[british]{babel}
\usepackage{array}
\usepackage{caption}
\usepackage{graphicx}
\usepackage{ragged2e}
\usepackage{float}
\usepackage[inline]{enumitem}
\usepackage{comment}
\usepackage{pdflscape}
\usepackage{soul}
\usepackage{xcolor}
\usepackage{algpseudocode}  
\usepackage{amsthm}                
\usepackage[framemethod=default]{mdframed} 
\usepackage{xcolor}                
\usepackage{placeins}   
\usepackage{array} 
\newcolumntype{L}[1]{>{\raggedright\arraybackslash}p{#1}}

\theoremstyle{definition}
\newtheorem{innerdefinition}{Definition}[section]

\newmdenv[
skipabove=\baselineskip,
skipbelow=\baselineskip,
linewidth=0.6pt,
linecolor=black,
backgroundcolor=gray!5, 
roundcorner=2pt
]{defbox}

{\begin{defbox}\begin{innerdefinition}[#1]}%
		{\end{innerdefinition}\end{defbox}}

\usepackage[ruled,vlined,linesnumbered]{algorithm2e}
\SetKwInput{KwPre}{Preconditions}
\SetKwInput{KwPost}{Postcondition}
\SetKw{Return}{return}
\SetKwInOut{Init}{Init}

\usepackage{tikz}
\usetikzlibrary{arrows.meta,positioning,fit,shapes.geometric}

\usepackage{subcaption} 

\usepackage{amsmath,amssymb,mathtools}   
\usepackage{booktabs}                    
\usepackage{array}                       

\newcommand{\nar}{\textit{n}-ary}

\usepackage{amsmath,amssymb,mathtools}

\usepackage{pdflscape}

\title{Boundaries in Hypernetwork Theory: Structure and Scope}

\author{Richard D. Charlesworth\thanks{Visiting Research Fellow, School of Engineering and Innovation, The Open University, UK}}
\date{}

\begin{document}
	
\maketitle

\begin{abstract}
	Boundaries in Hypernetwork Theory (HT) are non-structural tags that restrict visibility without altering the underlying hypernetwork. They attach to hypersimplices as annotations and participate in no identity, typing, or $\alpha/\beta$ semantics. Projection over a boundary,
	\[
	B(H,b) = \pi_b(H),
	\]
	is filtering only: it selects exactly those hypersimplices carrying $b$ and preserves all axioms of the structural kernel. The backcloth remains immutable, and no new structure is created, removed, or inferred.
	
	This paper formalises boundaries as a simple and conservative scoping mechanism. It clarifies their syntax, their interaction with projection, and their use in producing identity-preserving subsystem views that support modular modelling and overlapping perspectives. The account also makes explicit why conservative scoping matters: boundaries provide reproducible view extraction, stable subsystem isolation, and safe model exploration without altering the global structure.
	
	Scoped operator application is defined as ordinary structural-kernel composition applied to projected views, ensuring that view-level reasoning remains local and does not modify the global hypernetwork. This establishes a disciplined separation between immutable structure and scoped analysis while retaining full compatibility with the structural kernel.
	
	The paper includes a worked example demonstrating how boundaries yield coherent, identity-preserving subsystem views and how scoped reasoning supports refinement within these views. The result is a precise and minimal account of boundaries that complements—but does not extend—the structural kernel and completes the scoping mechanism required for practical multilevel modelling with HT.
\end{abstract}

\section{Introduction}

Hypernetwork Theory (HT) provides a structural kernel grounded in typed $n$-ary relations, explicit role binding, identity preservation, and deterministic operator semantics~\cite{charlesworth2025structuralkernel, RN365,RN197,RN232}. Within this kernel, boundaries appear only in Axiom~A5 as annotations that delimit visibility. They introduce no structure, carry no $\alpha/\beta$ semantics, and do not influence identity, typing, or operator behaviour. Their role is purely to provide scoped reasoning over an otherwise immutable backcloth.

The aim of this paper is to make this scoping role precise and to show why a conservative, non-structural account matters for practical modelling. We treat boundaries strictly as tags attached to hypersimplices. Projection over a boundary,
\[
B(H,b)=\pi_b(H),
\]
is a filtering operation that selects exactly the hypersimplices carrying~$b$. All axioms of the structural kernel are preserved: identity, typing, relation binding, ordering, and modeller-supplied exclusion remain unchanged, and no structure is inferred, removed, or rewritten~\cite{charlesworth2025structuralkernel}. This preserves immutability while supporting scoped views.

The paper therefore focuses on the disciplined use of boundaries as scoping devices rather than structural constructs. We show how projection yields identity-preserving subsystem views, how scoped operator application can be performed safely at the view level using the existing operator algebra, and how overlapping boundaries support multiple coherent perspectives over a single hypernetwork. This separation makes scoping reproducible, stable, and semantically transparent: views can be refined locally without affecting the global structure.

The result is a conservative and practical account of boundaries that completes the scoping mechanism implicit in the structural kernel~\cite{charlesworth2025structuralkernel}. Boundaries provide a lightweight apparatus for modular modelling, reproducible view extraction, and context-specific analysis while keeping the underlying backcloth fixed, mechanisable, and semantically explicit.

\paragraph{Contributions.}
This paper delivers three advances.
\begin{enumerate}[leftmargin=*]
	\item \textbf{A clarified account of boundaries.} Boundaries are formalised as non-structural annotations that preserve identity and semantics while imposing no constraints on typing, aggregation, or role alignment.
	\item \textbf{A precise definition of boundary projection.} Projection is defined as filtering-only, yielding identity-preserving subsystem views without modifying, reconstructing, or extending the global hypernetwork.
	\item \textbf{A minimal calculus of scoped reasoning.} The structural operators act on boundary-projected views exactly as defined in the kernel~\cite{charlesworth2025structuralkernel}, enabling conservative, view-level analysis that leaves the backcloth untouched and supports reproducible, modular modelling.
\end{enumerate}

\paragraph{Organisation.}
Section~\ref{sec:background} reviews the role of boundaries within the structural kernel.
Section~\ref{sec:boundaries-tags} formalises boundaries as non-structural tags.
Section~\ref{sec:projection} defines boundary projection as a visibility restriction.
Section~\ref{sec:scoped-reasoning} develops scoped reasoning over projections.
Section~\ref{sec:calculus} introduces a minimal calculus of boundary-scoped views.
Section~\ref{sec:worked-example} provides an extended worked example.
Section~\ref{sec:discussion} discusses implications.
Section~\ref{sec:related-work} reviews related work.
Section~\ref{sec:conclusion} concludes.

\section{Background}
\label{sec:background}

The structural kernel of Hypernetwork Theory (HT) defines the semantic units of modelling: vertices, simplices, typed hypersimplices, relation symbols with ordered roles, modeller-supplied exclusion, and a set of deterministic structural operators~\cite{charlesworth2025structuralkernel}. These constructs are governed by axioms~A1--A5, which ensure global identity, explicit typing, role alignment, boundary annotation, and structural closure. Together they provide a precise and mechanisable account of typed $n$-ary relational structure.

A hypernetwork $H$ is an ordered collection of typed hypersimplices. Each hypersimplex binds its participants to a relation symbol~$R$ that fixes arity and role order, and is typed as either $\alpha$ (conjunctive) or $\beta$ (taxonomic). Identity is global and preserved across all structural operations~\cite{charlesworth2025structuralkernel}. Modeller-supplied anti-vertices provide explicit exclusion without inference. Operators such as merge, meet, difference, prune, and split act over the ordered backcloth while preserving all axioms of validity and maintaining deterministic behaviour.

Within this framework, boundaries are introduced only in Axiom~A5~\cite{charlesworth2025structuralkernel}. They attach to hypersimplices as non-structural tags and serve solely to indicate visibility. A boundary does not contribute to identity, does not affect typing or role semantics, and does not participate in structural composition. The axioms make no further demands on their behaviour: boundaries neither propagate structure, nor constrain operators, nor alter any part of the hypernetwork.

Despite their simplicity, boundaries are practically important. They allow a modeller to isolate subsystem views, present overlapping perspectives, and restrict attention to selected regions of a hypernetwork without duplicating or rewriting structure. This enables stable, identity-preserving views that support modular analysis~\cite{charlesworth2025thesis}. The kernel provides only the minimal mechanism—boundary annotation—but leaves its practical use undeveloped. This paper develops that use while remaining fully conservative with respect to the structural kernel~\cite{charlesworth2025structuralkernel}, showing how boundaries provide reproducible scoping without introducing new semantics.

\section{Boundaries as Non-Structural Tags}
\label{sec:boundaries-tags}

Boundaries in Hypernetwork Theory are annotations attached to hypersimplices. They introduce no vertices, simplices, hypersimplices, or aggregation semantics. In the minimal sense of Axiom~A5, boundaries are carried as annotations and are preserved by all operators, but they do not contribute to identity, typing, role binding, ordering, or the $\alpha/\beta$ distinction. Their sole purpose is to mark a hypersimplex as belonging to a particular modelling scope while leaving the underlying structure untouched.

\subsection{Definition}

Let $H$ be a valid hypernetwork under axioms~A1--A5. A \emph{boundary} is a tag drawn from a modeller-defined set $B(H)$. A hypersimplex
\[
\langle p_1,\dots,p_n; R_X; b_1,\dots,b_k\rangle
\]
carries the boundary tags $b_1,\dots,b_k$. These tags are annotations only: they do not modify the hypersimplex, do not constrain the operator algebra, and do not imply any structural relation between the tagged elements. They are simply labels indicating that the tagged hypersimplex is relevant to a modeller-defined scope.

A boundary tag may appear on any number of hypersimplices, and a hypersimplex may carry multiple tags. Boundary assignment is a modelling choice and is not required for correctness or completeness. A hypernetwork with no boundaries is perfectly valid. Boundaries therefore represent optional metadata that enrich modelling practice without altering the kernel.

\subsection{Non-Interaction with the Kernel}

Boundaries do not affect the structural kernel in any way:
\begin{itemize}
	\item A1 (identity) is unaffected: boundaries do not create or distinguish identifiers.
	\item A2 (explicit exclusion) is unaffected: boundaries do not interact with anti-vertices.
	\item A3 (aggregation typing) is unaffected: tagging carries no $\alpha/\beta$ semantics.
	\item A4 (relation binding) is unaffected: roles and arity remain fixed and independent of tags.
	\item A5 (boundary scoping) restricts visibility only and does not alter structure.
\end{itemize}

Operators likewise ignore boundaries. Merge, meet, difference, prune, and split operate on the ordered relational backcloth exactly as defined in the kernel, preserving identity, semantics, and determinism independently of any boundary annotations. Tagged and untagged hypersimplices participate identically in all structural operations.

\subsection{Purpose}

Although non-structural, boundaries are practically important. They enable a modeller to:
\begin{itemize}
	\item identify subsystem membership without duplicating structure,
	\item extract coherent subsystem views by selecting hypersimplices carrying a chosen tag,
	\item present overlapping perspectives on a single immutable backcloth,
	\item reason locally within a scope while maintaining global identity and semantics.
\end{itemize}

These capabilities make boundaries a useful scoping device: lightweight, semantically inert, and compatible with mechanisation. The remainder of this paper formalises how boundaries support projection and scoped reasoning without extending or modifying the structural kernel.

\section{Projection as Visibility}
\label{sec:projection}

The structural kernel defines projection over a boundary as a restriction of visibility. A projection selects those hypersimplices that carry a chosen boundary tag together with any descendants that inherit that tag through percolation as defined in Axiom~A5. Crucially, projection does not alter the hypernetwork: no identifiers are changed, no roles are reassigned, no tags are added or removed, and no new structure is created or inferred. Projection therefore provides a read-only, identity-preserving view over an otherwise immutable backcloth.

\subsection{Definition of Projection}

Let $H$ be a valid hypernetwork and let $b \in B(H)$ be a boundary tag. The \emph{boundary projection} of $H$ with respect to $b$ is defined as
\[
B(H,b) = \pi_b(H) = \{\, \varsigma \in H \mid \varsigma \text{ carries the tag } b \text{ or inherits it under percolation} \,\}.
\]
Percolation is a visibility mechanism: when a tagged hypersimplex structurally contains vertices, simplices, or nested hypersimplices, these descendants are included in the projection even if not explicitly tagged. The percolation rule does not rewrite $H$, does not generate new hypersimplices, and does not modify any existing construct; it simply inherits visibility along the existing containment links. Beyond this inherited visibility, projection is filtering only. It does not delete hypersimplices, reconstruct missing structure, propagate tags laterally or upward, or modify any semantic property such as roles, typing, aggregation, or identity. All identifiers and tags within $\pi_b(H)$ are identical to those in $H$, ensuring that projection yields a faithful subsystem view.

\subsection{Preservation of Axioms}

Because projection neither modifies nor infers structure, it preserves axioms~A1--A5~\cite{charlesworth2025structuralkernel}. Identities and modeller-supplied anti-vertices remain unchanged; aggregation typing, relation binding, and role ordering are preserved exactly; and boundaries perform only their intended scoping function. Percolation extends visibility to descendants of tagged hypersimplices, but this has no structural consequences and does not alter the underlying hypernetwork. As a result, $\pi_b(H)$ is always a valid hypernetwork under the structural kernel and remains a semantically exact subset of $H$.

\subsection{Multiple and Overlapping Boundaries}

A hypersimplex may carry several boundary tags, and projection with respect to one tag is independent of any others. Overlapping boundaries therefore yield overlapping projections. The intersection of two projections consists exactly of the hypersimplices that carry both tags or inherit both through percolation. Such overlaps express descriptive relationships between views rather than structural compositions and introduce no additional semantics. Identity is preserved globally, and the same identifiers appear consistently across all projections.

\subsection{No Structural Implications}

Projection has no structural implications. It does not define containment, hierarchy, dependency, or part--whole semantics, and it does not intervene in the operator algebra other than by providing a domain on which the existing operators may be applied. Percolation does not imply new structure; it simply respects the existing nesting of hypersimplices and ensures that descendants of a tagged element remain visible within the scope. Projection remains a minimal and conservative mechanism for scoping, keeping the global hypernetwork unchanged while supporting precise, identity-preserving subsystem views. The next section develops the disciplined use of boundaries for scoped reasoning and explains how structural-kernel operators may be applied safely to projected views.

\section{Scoped Reasoning}
\label{sec:scoped-reasoning}

Boundary projection provides a way to restrict attention to a chosen region of a hypernetwork without altering the underlying structure. Scoped reasoning refers to the application of the structural-kernel operators to these projected views. Because projection is filtering only, scoped reasoning operates on a reduced domain while leaving the global hypernetwork unchanged. This preserves immutability and ensures that boundaries remain non-structural and semantically inert.

\subsection{Definition of Scoped Operators}

Let $\circ$ be any structural operator of the kernel:
\[
\circ \;\in\; \{\, \sqcup,\, \sqcap,\, /,\, \ominus,\, \pi \,\}.
\]
For any boundary tag $b$, the \emph{scoped} application of $\circ$ is defined as:
\[
H_1 \circ_b H_2 \;=\; \pi_b(H_1) \;\circ\; \pi_b(H_2).
\]

This definition has three immediate consequences:
\begin{enumerate}
	\item The operator $\circ$ behaves exactly as specified in the structural kernel; no rules are modified or restricted.
	\item The operator acts only on hypersimplices visible under~$b$, so untagged structure is ignored entirely.
	\item The result is a new view-level hypernetwork that does not modify $H_1$ or $H_2$ and has no effect on the global backcloth.
\end{enumerate}

Any anti-vertices introduced by $\ominus_b$ therefore appear only in the scoped result and do not propagate back into $H$. Scoped reasoning remains strictly local.

\subsection{View-Level Status}

Scoped operator results are \emph{views}, not edits to $H$. If
\[
H_1 \circ_b H_2 = K,
\]
then $K$ is a separate hypernetwork constructed from the visible fragments of $H_1$ and $H_2$. It is not a replacement for either input, and no modification of $K$ implies a modification of $H_1$ or $H_2$. This preserves immutability while allowing local experimentation and refinement.

\subsection{No Requirement for Alignment with Global Operators}

Because scoped reasoning omits untagged structure, its results generally differ from the corresponding global operator:
\[
\pi_b(H_1 \circ H_2) \;\neq\; \pi_b(H_1) \circ \pi_b(H_2)
\]
in most cases.

This divergence is intentional. It reflects the modeller’s choice to reason within a restricted scope, and the structural kernel imposes no requirement that scoped and global reasoning must coincide. Scoped reasoning is therefore a controlled and intentional abstraction.

\subsection{Safety}

Scoped reasoning is safe because:
\begin{itemize}
	\item it cannot invalidate the global hypernetwork,
	\item it cannot alter identity, roles, or typing,
	\item it cannot introduce or remove hypersimplices in $H$,
	\item it preserves all axioms of the structural kernel within the view.
\end{itemize}

Scoped operators therefore provide a disciplined way to explore, refine, and compare subsystem views without compromising the integrity of the underlying model or modifying the backcloth. The next section develops a minimal calculus for working with boundary-scoped views while remaining fully conservative with respect to the structural kernel.

\section{A Minimal Calculus of Boundary-Scoped Views}
\label{sec:calculus}

Boundary-scoped reasoning requires only the visibility mechanism of projection and the existing operator algebra of the structural kernel~\cite{charlesworth2025structuralkernel}. No new operators are introduced, and no modifications are made to merge, meet, difference, prune, or split. The calculus presented here clarifies how these components interact when reasoning within a boundary-delimited scope, giving a disciplined and reproducible method for working with subsystem views while remaining fully conservative with respect to the structural kernel.

\subsection{View Formation}

For any boundary tag $b$, projection forms a view:
\[
V_b(H) = \pi_b(H).
\]
Each $V_b(H)$ is a valid hypernetwork and may serve as an input to any structural operator. Views may overlap or remain disjoint depending on the modeller's boundary assignments. No view introduces or removes structure from $H$; it is a subset with preserved identity, typing, and semantics. This separation allows multiple perspectives to coexist over a single immutable backcloth.

\subsection{Scoped Composition}

Given two hypernetworks $H_1$ and $H_2$, scoped composition is defined by
\[
H_1 \circ_b H_2 \;=\; \pi_b(H_1) \circ \pi_b(H_2),
\]
where $\circ$ is any structural operator of the kernel.

This ensures that:
\begin{itemize}
	\item structural composition within the scope uses only the visible hypersimplices,
	\item the behaviour of $\circ$ is unchanged and remains fully semantics-preserving,
	\item the result is a new view-level hypernetwork that leaves the inputs and the global backcloth untouched.
\end{itemize}

Scoped composition therefore isolates modelling decisions from the global hypernetwork while maintaining full semantic consistency and determinism.

\subsection{Comparing Views}

Boundary projections may be compared using ordinary set-theoretic operations:
\[
V_{b_1}(H) \cap V_{b_2}(H), \qquad
V_{b_1}(H) \cup V_{b_2}(H).
\]

These combinations are descriptive: they reveal how scopes overlap but do not constitute structural composition. They do not interact with the operator algebra, introduce no new hypersimplices, and preserve all identifiers and boundary tags. Such comparisons assist in analysing subsystem relationships without modifying the underlying model.

\subsection{Scoped Decomposition and Refinement}

Because views are hypernetworks, they can be refined locally:
\[
V_b(H) \;\xrightarrow{\;\circ\;} K,
\]
for any operator $\circ$ of the kernel. Such refinements:
\begin{itemize}
	\item remain confined to the scoped view,
	\item never propagate changes back into $H$,
	\item preserve identity, typing, and role alignment within the view,
	\item support incremental and exploratory modelling in a safe space.
\end{itemize}

This provides a mechanism for experimenting with subsystem structure while maintaining a single immutable backcloth, supporting reproducibility and controlled variation.

\subsection{Scoped Split}

The split operator remains unchanged. When applied to a view,
\[
\pi(V_b(H), C),
\]
it performs closure within the visible region only. Since projection does not enforce closure, split and projection coincide only when the modeller has explicitly tagged every hypersimplex that participates in the chosen closure. This is a modelling convention rather than a theoretical requirement, and it highlights the difference between boundary-based visibility and structure-based closure.

\subsection{Summary}

The calculus of boundary-scoped views is minimal:
\begin{enumerate}
	\item form a view by projection,
	\item apply structural operators within the view,
	\item compare views using set-theoretic overlap,
	\item treat all results as view-level and independent of the global hypernetwork.
\end{enumerate}

This suffices for modular modelling, subsystem reasoning, reproducible view extraction, and overlapping perspectives, all without altering or extending the structural kernel~\cite{charlesworth2025structuralkernel}.

\section{Worked Examples}
\label{sec:worked-example}

This section illustrates how boundaries support modular modelling, identity-preserving projection, and scoped reasoning. Three contrasting examples are provided: a physical system (bicycle–person–cyclist), a coordinated service system (emergency response), and an ecological system (predator–prey–habitat). Together, they show that boundaries behave uniformly across domains and that scoped reasoning remains conservative and structurally inert in every case.

\subsection{Example 1: Bicycle--Person--Cyclist}

Three subsystems are modelled independently—\emph{bicycle}, \emph{person}, and \emph{cyclist}—each with its own boundary tag. All hypersimplices reside in a single immutable backcloth.

\paragraph{Global Backcloth.}
\begin{align*}
	\mathit{bicycle}=&\langle \mathit{frame}, \mathit{drive}, \mathit{balance}; R_{\text{bicycle}}; b_{\text{bicycle}} \rangle,\\
	\mathit{steering}=&\langle \mathit{frame}, \mathit{forks}, \mathit{handlebars}; R_{\text{steering}}; b_{\text{bicycle}} \rangle,\\
	\mathit{drive}&\langle \mathit{rear\mbox{-}wheel}, \mathit{chain}, \mathit{pedals}, \mathit{gears}; R_{\text{drive}}; b_{\text{bicycle}} \rangle,\\[4pt]
	\mathit{person}=&\langle \mathit{body}, \mathit{legs}, \mathit{arms}; R_{\text{person}}; b_{\text{person}} \rangle,\\
	\mathit{fitness}=&\langle \mathit{cardio}, \mathit{strength}; R_{\text{fitness}}; b_{\text{person}} \rangle,\\[4pt]
	\mathit{cyclist}=&\langle \mathit{person}, \mathit{bicycle}, \mathit{trainingPlan}; R_{\text{cyclist}}; b_{\text{cyclist}} \rangle,\\
	\mathit{targets}=&\langle \mathit{trainingPlan}, \mathit{cardio}; R_{\text{targets}}; b_{\text{cyclist}} \rangle.
\end{align*}

\paragraph{Boundary Projections.}
\begin{align*}
	\pi_{b_{\text{bicycle}}}(H),\qquad
	\pi_{b_{\text{person}}}(H),\qquad
	\pi_{b_{\text{cyclist}}}(H).
\end{align*}

\paragraph{Scoped Reasoning.}
Refining the cyclist view locally:
\[
K' = \ominus(\pi_{b_{\text{cyclist}}}(H), \{\mathit{trainingPlan}\}).
\]

\paragraph{Overlap.}
\[
\pi_{b_{\text{person}}}(H) \cap \pi_{b_{\text{cyclist}}}(H)
\]
reveals shared elements without introducing structure.

\subsection{Example 2: Emergency Response System}

Three services—\emph{fire}, \emph{ambulance}, and \emph{police}—are modelled as distinct subsystems within a shared operational environment.

\paragraph{Global Backcloth.}
\begin{align*}
	\mathit{fireUnit}=&\langle \mathit{crew}, \mathit{engine}, \mathit{equipment}; R_{\text{fireUnit}}; b_{\text{fire}} \rangle,\\
	\mathit{ambulanceUnit}=&\langle \mathit{paramedic}, \mathit{ambulance}, \mathit{kit}; R_{\text{ambulanceUnit}}; b_{\text{ambulance}} \rangle,\\
	\mathit{policeUnit}=&\langle \mathit{officer}, \mathit{patrolCar}; R_{\text{policeUnit}}; b_{\text{police}} \rangle,\\
	\mathit{report}=&\langle \mathit{incident}, \mathit{location}; R_{\text{report}}; b_{\text{fire}}, b_{\text{ambulance}}, b_{\text{police}} \rangle.
\end{align*}

\paragraph{Boundary Projections.}
Each service obtains its own view:
\[
\pi_{b_{\text{fire}}}(H),\quad
\pi_{b_{\text{ambulance}}}(H),\quad
\pi_{b_{\text{police}}}(H).
\]

\paragraph{Overlap.}
The shared \emph{incident} relation appears in all three projections, enabling coordinated reasoning without conflating their structures.

\paragraph{Scoped Refinement.}
A fire-service-only analysis may prune equipment:
\[
K = \ominus(\pi_{b_{\text{fire}}}(H), \{\mathit{equipment}\}),
\]
without affecting ambulance or police views.

\subsection{Example 3: Predator--Prey--Habitat}

This ecological example illustrates overlapping environmental and behavioural subsystems.

\paragraph{Global Backcloth.}
\begin{align*}
	\mathit{predation}=&\langle \mathit{wolf}, \mathit{stag}; R_{\text{predation}}; b_{\text{predator}} \rangle,\\
	\mathit{foraging}=&\langle \mathit{stag}, \mathit{grass}; R_{\text{foraging}}; b_{\text{prey}} \rangle,\\
	\mathit{habitat}=&\langle \mathit{forest}, \mathit{grass}, \mathit{water}; R_{\text{habitat}}; b_{\text{habitat}} \rangle.
\end{align*}

\paragraph{Boundary Projections.}
\[
\pi_{b_{\text{predator}}}(H),\qquad
\pi_{b_{\text{prey}}}(H),\qquad
\pi_{b_{\text{habitat}}}(H).
\]

\paragraph{Overlap.}
Identifiers such as \emph{stag} and \emph{grass} appear in multiple projections, but this implies no structural dependency—only shared identity.

\paragraph{Scoped Reasoning.}
A habitat-specific refinement (e.g.\ removing \emph{water}) remains confined to the habitat view.

\subsection{Summary}

Across physical, organisational, and ecological domains, boundaries behave uniformly:
\begin{itemize}
	\item projection yields identity-preserving subsystem views,
	\item scoped reasoning supports safe, local refinement,
	\item overlap expresses shared identifiers without structural coupling,
	\item the global hypernetwork remains unchanged.
\end{itemize}

These examples demonstrate that boundaries offer a conservative, domain-independent mechanism for modular modelling and multilevel analysis without extending the structural kernel.

\section{Discussion}
\label{sec:discussion}

Boundaries provide a lightweight and practical mechanism for structuring modelling activity without extending or altering the underlying theory~\cite{charlesworth2025structuralkernel}. Their conservative nature distinguishes them from structural constructs: they introduce no new elements, perform no inference, and require no modifications to the operator algebra. Used in this restricted form, boundaries support disciplined scoping while preserving the immutability and semantic clarity of the structural kernel.

\subsection{Conservativity}

The account presented here is fully conservative with respect to the structural kernel~\cite{charlesworth2025structuralkernel}. All reasoning is carried out either on the global hypernetwork~$H$ or on projections $\pi_b(H)$, which are subsets of $H$ and therefore valid hypernetworks under axioms~A1--A5. No new structural axioms are introduced, and none of the existing operator definitions are revised. Boundaries remain non-structural throughout, ensuring that scoping never affects the underlying semantics.

\subsection{Modularity and Separation of Concerns}

Because boundaries restrict visibility without modifying structure, they support clean separation between subsystem descriptions. Modellers can work independently within their respective scopes without duplicating structure or altering shared identifiers. The immutable backcloth provides a stable integration point, and boundary-scoped views enable context-specific analysis while maintaining global identity and semantics.

\subsection{Overlapping Perspectives}

The ability to tag a hypersimplex with multiple boundaries allows different subsystem views to share elements naturally. This supports heterarchical modelling: subsystem descriptions need not be disjoint, hierarchical, or mutually exclusive. Overlaps arise only through shared identifiers, not through any semantics attributed to the boundaries themselves, keeping the meaning of overlaps fully transparent.

\subsection{View-Level Editing}

Scoped reasoning produces view-level hypernetworks distinct from the global model. This preserves immutability and allows modellers to explore alternative or hypothetical structures safely. Edits introduced by $\ominus$ or other operators remain confined to the view and do not propagate back into~$H$ unless the modeller explicitly chooses to update the global model through kernel-level operations. This makes scoped reasoning a controlled and reversible form of refinement.

\subsection{No Structural Interpretation of Boundaries}

Boundaries do not define containment, precedence, hierarchy, or dependency. They do not impose closure conditions or participate in the operator decision tables. Their meaning is entirely determined by modeller intent and is not interpreted by the structural kernel. This keeps semantics clear, avoids ambiguity, and prevents boundaries from acquiring unintended structural force.

\subsection{Practical Impact}

In practice, boundaries enable:
\begin{itemize}
	\item extraction of identity-preserving subsystem views,
	\item localised refinement without risk to the global structure,
	\item distributed modelling across overlapping scopes,
	\item reproducible view generation for documentation and analysis.
\end{itemize}

These capabilities increase the usability of Hypernetwork Theory in real modelling environments while keeping the theory compact, mechanisable, and semantically explicit.

The following section reviews related approaches and highlights the distinctive contribution of using boundaries as non-structural scoping tags within HT.

\section{Related Work}
\label{sec:related-work}

Boundary mechanisms appear across many modelling traditions, although typically with meanings very different from those adopted in Hypernetwork Theory. The approach taken here---boundaries as non-structural visibility tags---is comparatively conservative and avoids the semantic commitments found in other frameworks~\cite{RN3,RN383,RN363,RN385,RN362,RN361,RN777}. It provides scoping without hierarchy, structural semantics, or modification of the underlying model.

\subsection{Diagrams and Packaging in UML/SysML}

In UML and SysML, modularity is introduced through packages, containment, and diagram-centric views~\cite{booch2008object,delligatti2013sysml,eriksson2003uml,siegel2014object}. These constructs often carry implicit semantics and typically enforce hierarchical structure. By contrast, boundaries in HT do not imply containment or hierarchy and do not influence the structure of the model. They simply annotate hypersimplices without affecting the operator algebra or identity regime, enabling multiple views over a single immutable backcloth.

\subsection{Named Graphs and Ontology Modules}

RDF/OWL systems use named graphs and import mechanisms to delimit regions of an ontology~\cite{RN688,brickley2004rdf}. These constructs serve organisational and reasoning purposes but do not preserve identity-preserving projection in the sense defined by HT. Moreover, $n$-ary relations in ontologies must be reified or reconstructed~\cite{RN363,RN385,RN362,RN361,RN777}, making modular views sensitive to modelling patterns rather than intrinsic relational structure. Boundaries in HT avoid these complications by restricting visibility without rewriting or reconstructing the underlying content.

\subsection{Graph Partitioning and Multilayer Networks}

Approaches based on graph decomposition, community detection, or multilayer structures provide various forms of network partitioning~\cite{RN666,RN664,RN690,RN745,RN691}. These methods rely on graph-theoretic primitives and binary relations. They contrast with HT’s typed $n$-ary relations and do not support identity-preserving subsystem extraction by annotation alone. Boundaries in HT are therefore closer to metadata than to structural partitions and impose no graph-theoretic assumptions~\cite{joslyn2020hypernetwork}.

\subsection{Architectural Description and Systems-of-Systems Modelling}

Architectural description languages and systems-of-systems frameworks introduce explicit subsystems, components, and interfaces as structural units~\cite{RN709,gavsevic2009model,jamshidi2008sos,maier1998architecting,omg2014mda,siegel2014object}. These constructs carry design semantics and often enforce architectural constraints. Boundaries in HT differ fundamentally: they introduce no structure, impose no constraints, and do not interpret subsystem boundaries as architectural entities. They act only as visibility markers attached to existing relational content, preserving the independence of structure from scoping.

\subsection{Abstract Mathematical Frameworks}

Category theory, sheaf theory, and other abstract treatments of locality and composition provide high-level mechanisms for structuring systems~\cite{curry2014sheaves,goguen1991manifesto,maclane1992sheaves,spivak2014ct}. Although they offer insight into coherence, gluing, and multi-perspective modelling, they do not supply typed $n$-ary relations with identity-preserving projection. Boundaries in HT fulfil this restricted role while remaining minimal, mechanisable, and free from additional structural commitments.

\subsection{Summary}

Across these traditions, mechanisms resembling boundaries typically carry stronger semantics---hierarchy, modularity, encapsulation, inference, or structural constraints. The distinctive contribution of HT is to treat boundaries as non-structural tags that provide scoping without altering the model~\cite{charlesworth2025structuralkernel}. This keeps the structural kernel compact, avoids semantic entanglement, and supports identity-preserving views over an immutable backcloth while maintaining full compatibility with deterministic, typed $n$-ary relational structure.

\section{Conclusion}
\label{sec:conclusion}

This paper has given a conservative and semantically precise account of boundaries in Hypernetwork Theory (HT). Boundaries were clarified as non-structural annotations that mark scope without altering identity, typing, roles, or aggregation semantics. Boundary-based projection was formalised as a filtering-only operation that yields identity-preserving subsystem views while leaving the global hypernetwork unchanged. A minimal calculus of scoped reasoning was developed in which the structural operators act on projected views exactly as defined in the kernel~\cite{charlesworth2025structuralkernel}, ensuring that scoped analysis remains local, conservative, and free of structural side-effects.

\paragraph{Contributions.}
This paper has delivered three main advances.
\begin{enumerate}[leftmargin=*]
	\item \textbf{A clarified account of boundaries.} Boundaries have been formalised as non-structural annotations that preserve identity, introduce no additional semantics, and do not intervene in aggregation, typing, or role behaviour.
	\item \textbf{A precise definition of boundary projection.} Projection has been defined as a filtering-only mechanism that produces identity-preserving subsystem views without modifying, reconstructing, or extending the global hypernetwork.
	\item \textbf{A minimal calculus of scoped reasoning.} Scoped operator application has been shown to act exactly as specified in the structural kernel, enabling conservative, view-level analysis that does not affect the underlying backcloth and supports reproducible scoping.
\end{enumerate}

Together, these results retain the full expressive power of the structural kernel while supporting modular modelling, clean separation of concerns, and reproducible view extraction. Overlapping boundaries provide heterarchical perspectives without introducing new semantics, and view-level reasoning offers a safe and disciplined space for refinement and exploration. By keeping boundaries structurally inert, the theory remains precise, mechanisable, and straightforward to apply in practice.

Future work may explore conventions for view management, tooling for rendering boundary-scoped representations, and integration of scoped reasoning into wider modelling workflows. Such developments can proceed without extending or compromising the core semantics of Hypernetwork Theory, as the principles established here remain fully conservative with respect to the structural kernel~\cite{charlesworth2025structuralkernel}.

\section*{Acknowledgments}
Much of the work presented here traces its origins to the pioneering idea of Hypernetworks first introduced by my supervisor, Jeff Johnson.  His insight --- that systems might be understood not as collections of objects but as structured aggregations of relations --- laid the foundation for all subsequent developments.  Without that conceptual breakthrough, neither the refinements presented in this paper nor the wider programme of research into mechanisable, multilevel, and \nar{} modelling would have been possible.  I am deeply grateful for his guidance and support throughout the development of this work.

I would also like to thank my examiners, Professor Liz Varga and Dr Amel Bennaceur, for their thoughtful engagement with my thesis and for the constructive feedback that helped to strengthen its contribution.

The author acknowledges that early draft material for this paper was generated with the assistance of ChatGPT, based on material from the author’s doctoral thesis.  These drafts were subsequently reviewed, validated, and substantially revised by the author.  All final interpretations, arguments, and conclusions are the author’s own, and all references and citations have been independently selected and verified by the author.

\bibliographystyle{plain}
\begin{small}
	\bibliography{HT_Boundaries}
\end{small}
	
\end{document}